\documentclass[12pt]{article}
\usepackage{graphicx}
\def\LL{Lorentz--Lorenz}
\def\LLc{Lorentz--Lorenz coefficient}

\def\LLf{Lorentz--Lorenz function}
\def\LLr{Lorentz--Lorenz relation}
\def\dfe{1,1-difluoroethylene}
\def\dfeform{${\rm H}_{2}{\rm C}_{2}{\rm F}_{2}$}
\def\cc{coexistence curve}
\def\ct{critical temperature}

\def\cp{critical phenomena}

\def\oneun#1{{\rm #1}}
\def\twoun#1#2{{\rm #1#2}}
\def\threeun#1#2#3{{\rm #1#2#3}}

\def\nm{{\rm nm}}

\def\cm{{\rm cm}}

\def\g{{\rm g}}

\def\K{{\rm K}}

\def\mol{{\rm mol}}

\long\def\symbolfootnote[#1]#2{\begingroup%
\def\thefootnote{\fnsymbol{footnote}}\footnote[#1]{#2}\endgroup}

\begin{document}
\begin{center}
    {\bf {\Large Lorentz--Lorenz Coefficient, Critical Point Constants,\\
and Coexistence Curve of 1,1-Difluoroethylene}}\\
\end{center}
\vspace{5 mm}

\begin{center}
Nicola Fameli\symbolfootnote[1]{Corresponding author, presently at : Department of 
Pharmacology and Therapeutics, The University of British
Columbia, 2176, Health Sciences Mall, Vancouver, B. C., Canada V6T 
1Z3; Electronic address: {\tt fameli@physics.ubc.ca}}, David A. Balzarini\\
{\it Department of Physics and Astronomy,} \\ 
{\it The University of British Columbia,}\\
{\it 6224 Agricultural Road, Vancouver, B. C., Canada V6T 1Z1}
\end{center}
\vspace{150mm}

%(\today)
\pagebreak

{\centerline{\Large Abstract}}
\vspace{5mm}

We report measurements of the \LLc\ density dependence, ${\cal
L}(\rho)$, the critical temperature, $T_{\rm C}$,
and the critical density, $\rho_{\rm C}$, of the fluid \dfe\
(\dfeform).
\LLc\ data were obtained by measuring refractive
index, $n$, and density, $\rho$, of the same fluid sample 
independently of one another.
Accurate determination of the \LLc\ is necessary for transformation of
refractive index data into density data 
from optics--based experiments on critical phenomena of fluid
systems done with different apparatus, with which independent
measurement of $n$ and $\rho$ is not possible.
Measurements were made along
the coexistence curve of the fluid and span the density range
$0.01\;{\rm to}\;0.80\;\g\,\cm^{-3}$. The \LLc\ results show a
stronger density
dependence along the \cc\ than previously observed in other fluids, 
with a monotonic decrease from a density of about
$0.2\;\g\,\cm^{-3}$ onwards, and an overall variation of about 2.5\%
in
the density range studied. No anomaly in the \LLf\ was observed 
near the critical density.
The critical temperature is measured at $T_{\rm C}=(302.964\pm
0.002)\;\K$ 
($29.814\;^{\circ}\oneun C$) and the measured critical density is
$\rho_{\rm C}=(0.4195\pm 0.0018)\;\g\,\cm^{-3}$.
\vspace{5mm}

%\pacs{60.64.Fr}


%\maketitle
\pagebreak

\section{\label{intro}Introduction}

We have accurately measured the density dependence of the
Lorentz-Lorenz
coefficient of 1,1-difluoroethylene (\dfeform), its 
critical temperature and critical density. 
The experiments presented herein constitute a preparatory phase to a
set of
other experiments we are carrying out with a different apparatus from 
the one described in this article. The latter set of experiments will
combine
three different optical techniques in one apparatus, for the study of
critical phenomena in pure fluids~\cite{pangsthesis, nfsthesis},
as well as the mapping of the $P$--$V$--$T$ space of a
fluid. Said optical
techniques produce measurements of
index of refraction of the fluid under study. However, to be able to
interpret 
experimental results on \cp\ in fluids obtained through this
apparatus and
to compare them with theoretical
predictions, refractive index data
have to be transformed into
density data, as density is the quantity
typically used by theories in their description of this class of 
\cp\ (for example, for the order parameter describing the 
coexistence curve temperature dependence)~\cite{fisher82}. As a
result, 
we need to measure the \LLc\ which is the purpose of the experiments
described below.

The substance investigated in the experiments presented herein
was chosen both for its relatively easily accessible critical point
and
because
of the limited amount of
accurate data on its critical region and its \LLc\
available in the literature. In general, we have noticed scarcity of
\LL\ and critical point data on hydrocarbons.

The \LLc, ${\cal L}$, relates the index of refraction, $n$, 
to the density of a gas, $\rho$,
in the following way:
\begin{equation}
{n^{2}-1\over
n^{2}+2}=\rho{\cal L}(n,\;\rho)\label{LL}
\end{equation}
This relation is the optical frequency
equivalent of the Clausius--Mossotti relation for
the dielectric constant, at lower frequencies of the electromagnetic
spectrum~\cite{bornwolf, jackson}. 
Equation~(\ref{LL}) is used in obtaining experimental data on fluid
density, from measurements of refractive 
index~(\cite{chae90, schmidt94, yata00, oleinikova00}).

One can relate refractive index to density using a $P$--$V$--$T$
curve,
for example, by measuring the refractive index as a function of
pressure and then relying on an equation of state to relate pressures 
to densities.
However, it is clearly more desirable not to rely on any
previous data to relate refractive index and density of the sample at 
hand, since such data are necessarily obtained with different samples,
containing, in general, 
different percentages, or types, of impurities. Correlation of
results from separate
experiments can often lead to inaccurate or incorrect conclusions,
especially in
experiments where we have diverging quantities and where precise
temperature control is necessary, as is the case near the critical
point. 
To avoid this problem, 
it necessary to have an accurate estimate of
the \LLc, with index of refraction and density measured on the same
sample, and independently of one another.

We have therefore carried out measurements of the \LLc\ of \dfe\
using a
 self--contained apparatus, which enables us to 
measure separately the index of refraction and the density of the
fluid
in question.

Early investigations of the Lorentz--Lorenz coefficient were carried
out on pure fluids and mixtures at low densities and generally showed
a linear increase with density~\cite{buckingham74}.  Some experiments
showed an
anomaly near the critical density. However, these experiments measured
refractive index only and required $P$--$V$--$T$ data from other
experiments
for analysis and interpretation~\cite{lucas65}. Our measurements of
$\cal L$ show a departure from both of these features.

\section{EXPERIMENTAL METHOD}\label{method}

The substance studied, \dfe\ (also known as vinylidene fluoride,
molecular weight
$64.035\;\g\,\mol^{-1}$), is a
colorless, flammable, non--toxic gas at room temperature and
atmospheric pressure. Its main use is in preparing polymers and
copolymers and as an intermediate in organic
synthesis~\cite{matheson}. The
material used in these experiments was purchased from Scott Specialty 
Gases and is of 99.4\% purity.

\begin{figure}\centering
\includegraphics[scale=0.4]{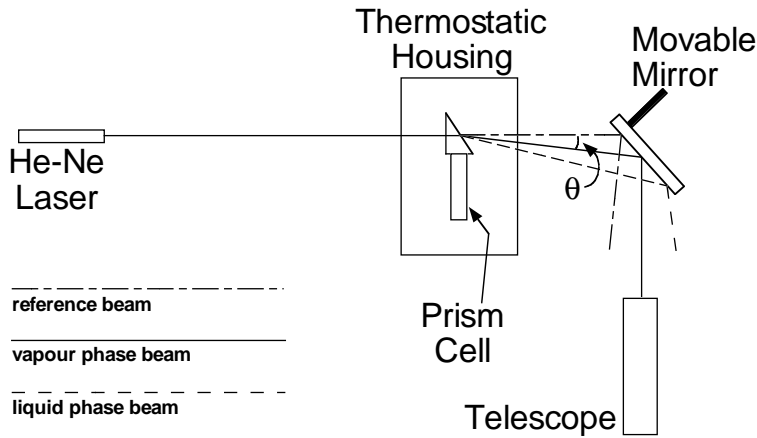}
\caption{Schematic diagram of the experimental apparatus 
(not to scale).}\label{prismcellexpt}
\end{figure}

In these experiments, the sample
is introduced in 
a high pressure container (the ``prism cell'' of
Fig.~\ref{prismcellexpt}), which has a prism--shaped section, about 
1~cm in height, formed
between two sapphire windows at one end of the cell~\cite{dbpp74}.
Sample temperature is
maintained at a chosen value by inserting the sample cell in a
thermostat capable of regulating temperature within $0.5\;\twoun mK$.
The experimental arrangement is
shown schematically in Fig.~\ref{prismcellexpt}.

The sample cell is filled with fluid at a high density and weighed.
It is then placed in the thermostatic housing. 
We measure the angle of deviation, $\theta$, 
with respect to the reference beam, of a
laser beam (He--Ne laser, $\lambda=632.8\;\nm$)
traversing the prism--shaped fluid sample. The beam is
expanded to a diameter of about 2.5~cm and collimated to ensure 
coverage of the entire sample cross--section. With this setup, when 
both liquid and vapor are present in the cell, both phases are 
sampled at the same time.
The deviation angle depends on the refractive index of the fluid.
The micrometer screw on the adjustable mirror is
calibrated to relate the micrometer scale reading to the refraction
angle.

\begin{figure}\centering
\includegraphics[scale=0.50]{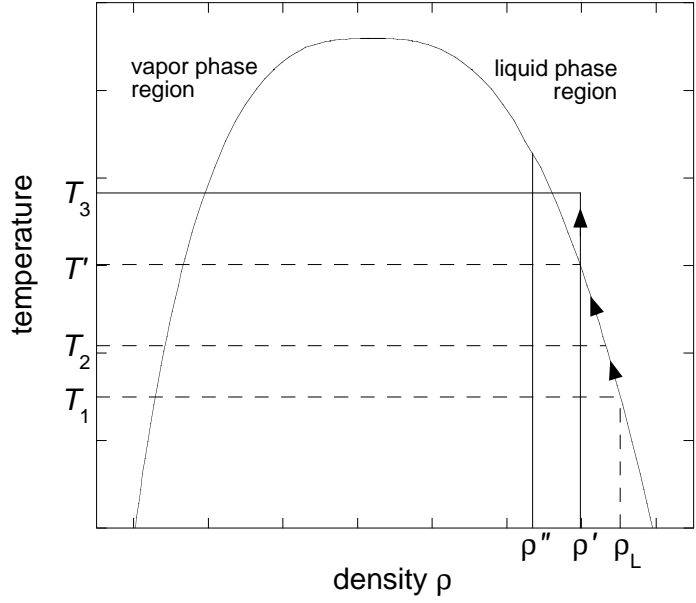}
\caption{Qualitative temperature--density 
phase diagram of the vapor coexistence
curve of \dfeform. This graph depicts the steps taken to
obtain the \LLc. The sample cell is initially filled to a  
density, $\rho'$, inserted in the 
thermostat set a temperature $T_{1}<T'$ and 
the liquid phase refractive index, $n_{L}(\rho',\,T_{1})$, is
measured. 
The temperature is then increased to $T_{2}$ and  
$n_{L}(\rho',\,T_{2})$ is measured and found to be different from 
$n_{L}(\rho',\,T_{1})$ as long as $T_{2}<T'$. The procedure is 
repeated until, for temperatures above $T'$, the measured 
$n_{L}(\rho',\,T)$ is practically independent of 
temperature. The system is then in the one phase region outside the 
coexistence curve and measurements of the refractive index 
in this situation yield $n_{L}(\rho',\,T')$ needed for the 
determination of $\cal L$ at density $\rho'$. 
The sample is then brought to 
another density $\rho''$ by bleeding out some fluid and the procedure 
repeated to obtain $n_{L}(\rho'',\,T'')$ and so
on.}\label{LLdatacollect}
\end{figure}

During the measurements, 
if the temperature is such that the fluid is below the coexistence
curve, both liquid and vapor phases are present.  Since the region of
interest is just outside the coexistence curve, the temperature is
gradually
increased until no further change is measured in the
angle of deviation.  This corresponds to having the fluid in a single
phase (Fig.~\ref{LLdatacollect} summarizes this procedure
graphically). 
The sample cell is then removed from the thermostatic housing
and weighed again.  This provides one datum of deviation angle
and sample cell mass.
Some fluid is bled from the sample, and the procedure is repeated to
provide another measurement.  This measurement sequence is repeated
until the
sample density is close to the critical value. At this stage of the
experiment, the coexistence curve is measured, by returning the sample
to the low temperature end of the measurement sequence and recording
the
deviation angle of the laser beam in the liquid and vapor phases as  
a function of sample temperature. Critical temperature and
critical density of the sample are then obtained from these data.

Following the coexistence curve determination, measurements
of mass and deviation angle are resumed until the sample cell is
empty, to complete the data set for the \LLc\ determination 
from the
critical density to near--zero density.
This results in a series of measurements of
deviation angle versus sample mass, corresponding to the region just
above the coexistence curve.  The reason
for choosing this region, in which to measure the \LLc, is
to be able to use this data in interpreting
measurements of the coexistence curve and critical point constants 
from this and other experiments.

The mass measurements, $m_{\rm cell+fluid}$, 
are of the sample fluid plus the sample
container masses, hence the mass of the fluid alone is obtained by 
subtracting
the mass of the empty cell, $m_{\rm cell}$, from the measured mass.
Then, having
measured the volume of the container, $V$, the sample density,
$\rho_{\rm
fluid}$, is obtained: $\rho_{\rm fluid}=(m_{\rm cell+fluid}-m_{\rm
cell})/V$.
The volume of the sample
container is determined 
by filling it with distilled water and weighing it. The volume of our 
experimental cell is $(12.066\pm0.003)\;{\rm cm}^{3}$, including a 
small correction accounting for volume change with temperature.

The index of refraction is obtained by optical analysis.
First, the adjustable mirror micrometer screw scale calibration
equation is determined. This is obtained by placing a diffraction
grating in the same position that the sample occupies during the
measurements and by taking readings on the micrometer scale of the 
diffraction maxima. These are, in turn, related to the refraction
angle, $\theta$, giving a linear
relationship between the micrometer scale and the beam
deflection angle~\cite{nfsthesis}. After this calibration stage,
application of Snell's law to our optical system leads to a
measurement of the refractive index of the fluid, $n_{\rm fluid}$, as 
a function of the refraction angle, $\theta$. 
Care must be exercised in this step of the analysis in
order to eliminate any effects of possible wedge angles in the
windows.  Any wedges between the cell window faces were measured
at the same time that the internal angle of the hollow prism was
measured, and were accounted for in our analysis~\cite{uli90}.
The angle of the hollow prism was also measured with high
pressure in the cell in order to take into account any significant
effects on the experimental measurements. The distortion of the volume
of the experimental cell with pressure is negligible in the pressure
range of the present experiment (up to about $15\;\threeun MPa$). 
Measurements of the index of refraction are taken across the whole 
sample cross--section. The accuracy of the refractive index 
determination near the critical point is limited by the strong 
density gradient caused by the large compressibility~\cite{dbpp74, 
uli90}. Therefore, near the critical point, the refractive index 
measurements are taken at a sufficiently high temperature (above 
$T_{\rm C}$) to avoid beam distortion.

\section{RESULTS AND DISCUSSION}

\subsection{\LLc}\label{LLsection}

\begin{figure}\centering
\includegraphics[scale=0.50]{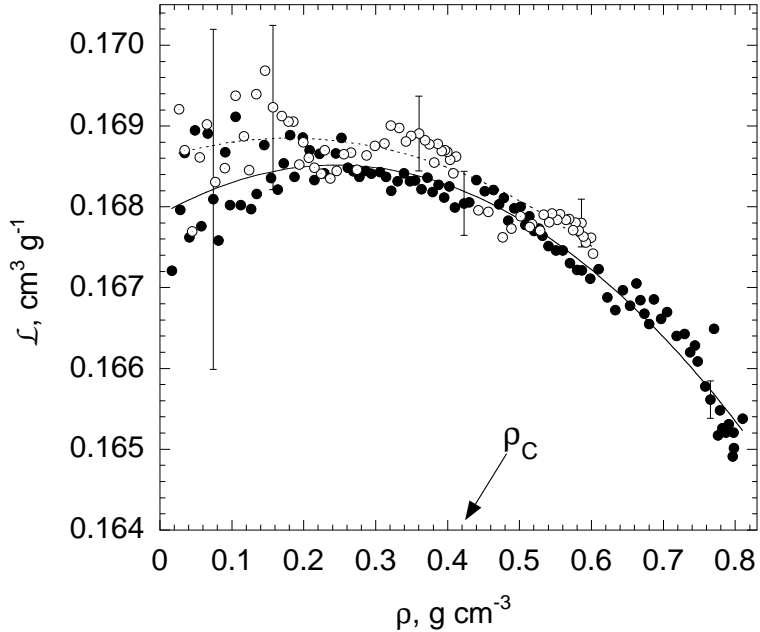}
\caption{Measured density dependence of the \LLc\
($\cal L$)
of 1,1-difluoroethylene. The solid and dotted curves are quadratic
fits to the 
data, according to Eq.~(\ref{virial}).
The fit parameters are in Table~\ref{LLfits}.}\label{LLdata}
\end{figure}

Results of the Lorentz--Lorenz coefficient measurements,
${\cal L}$, versus density are plotted in Fig.~\ref{LLdata}. The data
refer to 
two separate measurement sequences carried out during an
eight--month period, on two samples of \dfe\ extracted
from the same lecture bottle. One of the two
runs (black circles) extended to a higher density range than
the other (dotted circles). 

\begin{table*}
\begin{center}
    \begin{tabular}{llllc} \hline\hline
	& \hspace{3mm}${\cal L}_{0}\;(\cm^{3}\,\g^{-1})$ & 
	\hspace{3mm}${\cal L}_{1}\;(\cm^{6}\,\g^{-2})$ 
	& \hspace{3mm}${\cal L}_{2}\;(\cm^{9}\,\g^{-3})$ 
	& \hspace{3mm}${\cal L}_{\rm C}\;(\cm^{3}\,\g^{-1})$\\\hline
	run 1 (${\scriptstyle \odot}$) & \hspace{3mm}
	$0.1686\pm 0.0004$ & \hspace{3mm}$0.003\pm 0.003$ 
	& \hspace{3mm}$-0.008\pm 0.004$ & \hspace{3mm}$0.1685\pm 0.0017$\\
	run 2 ($\bullet$) & \hspace{3mm}
	$0.1679\pm 0.0003$ & \hspace{3mm}$0.005\pm 0.002$ 
	& \hspace{3mm}$-0.010\pm 0.002$ & \\\hline\hline
    \end{tabular}\label{LLfit}
    \caption{Results of a quadratic fit to the Lorentz--Lorenz data
of 
    1,1-difluoroethylene (C$_{2}$H$_{2}$F$_{2}$). ${\cal L}_{C}$ is
the
    critical \LLc\ calculated from the critical density $\rho_{\rm C}$
    (see procedure described in the next section.)}\label{LLfits}
\end{center}    
\end{table*}
The value of ${\cal
L}$ varies by approximately 2.5\% over the density range studied in
this
experiment exhibiting a dependence on the density along the \cc, 
with a gentle maximum in the neighborhood of
$\rho=0.2\;{\rm g}\,{\rm cm}^{-3}$.
Around the critical density, $\rho_{\rm C}$, where an accurate
determination of ${\cal L}$ is more crucial for our purposes,
the measured value of the \LLc\ varies by less than 1\% in both
sequences.

The \LLc\ can be expressed in a so--called refractometric virial
expansion in powers of the density~\cite{buckingham74}:
\begin{equation}
{\cal L}(\rho)={\cal L}_0+{\cal L}_1\rho+{\cal
L}_2\rho^2+...\label{virial}
\end{equation}
The solid and dotted lines shown in the graph of Fig.~\ref{LLdata}
represent quadratic
fits to the two series of data and yield the values of
the coefficients ${\cal L}_0$,
${\cal L}_1$, and  ${\cal L}_2$ reported in Table~\ref{LLfits}.  

Data at the low density end of the range investigated are affected 
by larger errors than the data around the critical density and larger 
density regions.
At low densities, the accuracy in the 
determination of $\cal L$ is mainly limited  by how accurately the 
mass of the empty prism cell can be measured at the end of the run.
As 
the quantity one needs is the difference $m_{\rm cell+fluid}-m_{\rm
cell}$, 
where $m_{\rm cell}$ is the mass of the empty cell, the same degree
of 
uncertainty in the empty cell mass yields a larger inaccuracy in the
determination of $\cal L$ at 
the low density than in the higher density measurements. Hence, the
larger scatter in the data in the low density region of the
measurements.

At the high density end of the
data, one must be aware of another experimental pitfall.  At those 
densities the liquid--vapor coexistence curve of 1,1-difluoroethylene
is at 
temperatures much lower than the typical room temperature, which was
monitored often during the experiment
and found to average around $23.3^{\circ}$C.  The 
reported data were taken starting at high densities, which 
meant maintaining the cell at temperatures of about 
$+3^{\circ}$C in the thermostat.
Clearly, then, when the cell is removed from the 
thermostatic housing to be weighed, condensation, and 
then evaporation, of atmospheric moisture on the cell body occurs
quite
rapidly thereby hindering accurate 
measurement of the cell mass. At the same time, at high densities, 
we could not afford to leave the cell out of the thermostat for very
long to wait for the condensation--evaporation effect to equilibrate,
since, in so doing, the entire sample would have quickly gone into the
liquid phase region of its phase diagram, rapidly reaching high 
enough pressures to cause possible damage to the experimental cell.
Instead,
to minimize this risk, a study 
of the consequences of the moisture 
condensation--evaporation phenomenon was done at the end of the
experiment, with the empty cell. It was found 
that when the  difference between the thermostat temperature and room
temperature was highest, the overall effect led to an underestimation
of about 0.5\% of the value of ${\cal L}$ corresponding to the region
at 
densities $\rho>0.75\;{\rm g}\,{\rm cm}^{-3}$. 
The monotonic decrease in 
the data in that density range is partially due to this effect.

Within experimental error, the data from the two
experimental runs overlap reasonably well and the slight
discrepancies between
the two sets can be ascribed to differences in the amount of
impurities in the samples. Previous studies of the effect of small
impurities on measurements of critical point constants have shown that
the critical temperature is much more sensitive to impurities than the
critical density is~\cite{hastings80, uli90}.

The estimated uncertainty in the $\cal L$ measurements is comparable
to the
scatter of data in the graph, as is illustrated by the error bars in
Fig.~\ref{LLdata}. It 
is about $2\times 10^{-4}\;{\rm
cm}^{3}\,{\rm g}^{-1}$ near the critical density.
This variation in $\cal L$ is
consistent with measurements made on other fluids in this
laboratory~\cite{dbpp74, burton74, uli90}. Moreover, as is apparent
from Fig.~\ref{LLdata}, we do not observe
any anomaly in the density dependence of the \LLc\ near the critical
point, within the limitations imposed 
on our optical technique by the strong density gradients occurring in 
pure fluids close to their 
critical point.
This result is 
in agreement with earlier theoretical studies on the
subject~\cite{lucas65, larsen65}.

From the \LL\ data, we also obtained the electronic (optical)
polarizability, $\alpha_{\rm p}$, of \dfe, using the
relationship $\lim_{\rho\rightarrow 0}{\cal
L}(\rho)=(4\pi\alpha_{\rm p}/3)N_{A}$, with $N_{A}$ 
Avogadro's number~\cite{buckingham74}:
the result is
$\alpha_{\rm p}=(4.29\pm0.01)$\AA$^{3}$.

\subsection{Coexistence curve, \ct, and critical 
density}\label{ccurve}

\begin{figure}\centering
\includegraphics[scale=0.6]{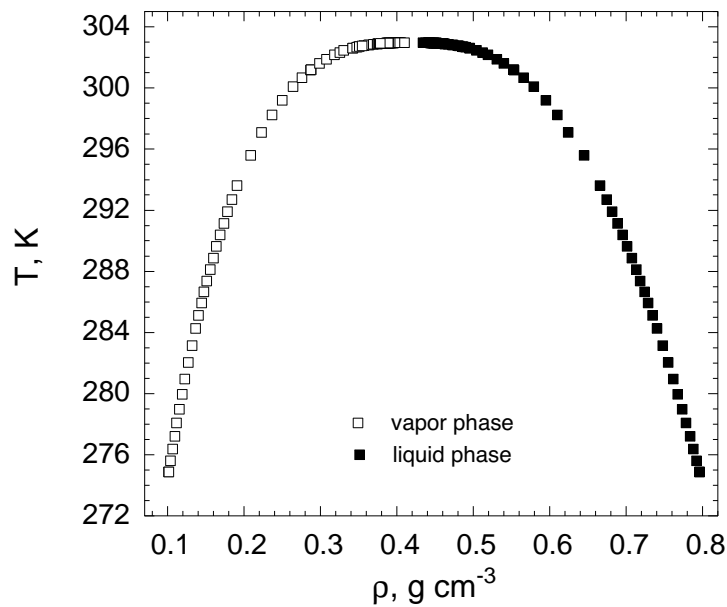}
\caption{Coexistence curve data obtained for \dfe; white squares 
correspond to vapor phase, black square to liquid phase.}\label{coexcurve}
\end{figure}

The liquid--vapor
coexistence curve of \dfe\ was also measured in this experiment.
This 
was done after the sample had been bled to the point corresponding as 
closely as possible to the
critical density.  The refractive index of the two coexisting phases
was measured as a function of temperature and then transformed into
a set of density measurements by means of the \LLc\ measured as 
described in section~\ref{LLsection}.
From these data, the critical temperature, $T_{\rm C}$, and the
critical density, $\rho_{\rm C}$, can be extracted. The coexistence 
curve data obtained in our experiments are shown in 
Fig.~\ref{coexcurve}.

The data have been analyzed in terms of renormalization group and
scaling theory~\cite{wilson71,wegner72,fisher82}, and
are fitted to an equation of the form,
\begin{equation}
\Delta\rho^{*}\equiv{\rho_{\rm L}-\rho_{\rm V}\over 2\rho_{\rm C}}=B_0
t^{\beta}\left( 1+B_1 t^{\Delta}+B_2 t^{2\Delta}+\cdots
\right),\label{corr2scaling}
\end{equation}
relating the order parameter $\Delta\rho^{*}$ to the reduced
temperature
$t=(1-T/T_{\rm C})$.
$\rho_{\rm C}$ is the critical density, $\beta$ the order parameter
critical exponent,
and $\Delta$ the correction--to--scaling critical exponent.

\begin{figure}\centering
\includegraphics[scale=0.5]{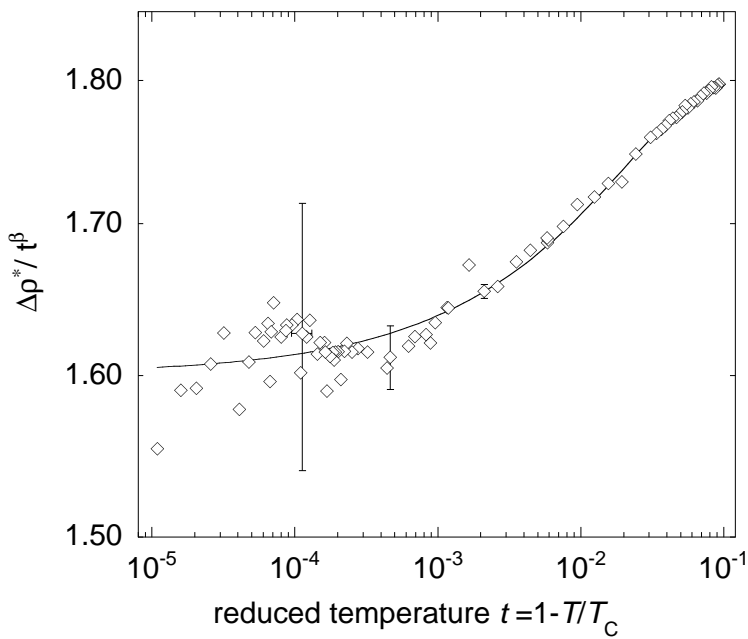}
\caption{$\log$--$\log$ plot of the order parameter data of \dfe. The
line represents data interpolation using Eq.~(\ref{corr2scaling}) over the range
$10^{-5}<t<10^{-1}$, with
$\beta=0.326$ and $\Delta=0.50$.}\label{coex}
\end{figure}

Fig.~\ref{coex} is a $\log$--$\log$ plot 
of $\Delta\rho^{*}/t^{\beta}$ as a function
of the reduced temperature, $t$, and it represents the
coexistence curve data according to 
Eq.~(\ref{corr2scaling}). The line interpolating the data
is a fit with two correction--to--scaling terms, with $\beta=0.326$
and 
$\Delta=0.50$ held fixed in the fits. The critical amplitude values, $B_{0}$, 
$B_{1}$, and $B_{2}$, obtained are as follows: $B_{0}=1.601\pm 
0.008$, $B_{1}=0.71\pm 0.06$, $B_{2}=-1.10\pm 0.16$.
The temperature range at which
coexistence curve data can be obtained is limited as the critical
point is approached because of ``gravitational rounding''
resulting from the increasing compressibility~\cite{moldover79}, as 
mentioned in section~\ref{method}.
In the reported measurements, 
the coexistence curve of C$_{2}$H$_{2}$F$_{2}$ was measured over the 
reduced temperature interval~$10^{-5}<t<10^{-1}$. 
However, the pronounced density gradients in the fluid nearing its
critical 
region render it arduous to take accurate data at values of 
$t{\mbox{\lower 1mm\hbox{$\stackrel{\textstyle <}{\sim}$}}}10^{-4}$,
as the much larger scatter in the data at low values of $t$
in Fig.~\ref{coex}
testifies.

\begin{figure}\centering
\includegraphics[scale=0.6]{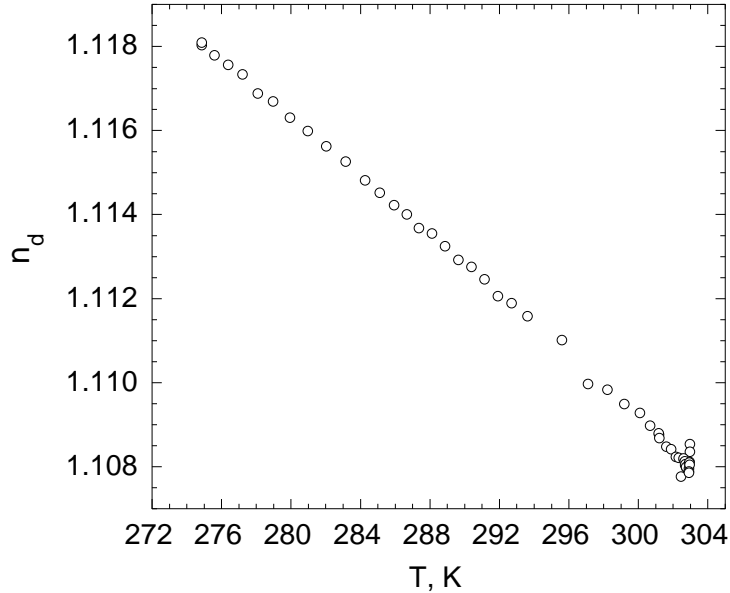}
\caption{Coexistence curve diameter data , $n_{\rm d}$, used to 
determine the critical refractive index, $n_{\rm C}$.}\label{diameter}
\end{figure}

In spite of this, we can still obtain an accurate estimate of the
critical density, $\rho_{\rm C}$, from the measured coexistence curve
data
by the following procedure.
First, the critical temperature, $T_{\rm
C}$, is obtained by fitting the coexistence curve 
data to a power law of the same form as Eq.~(\ref{corr2scaling}), but 
expressed in terms of the density difference, $\Delta\rho$, alone, not
divided by the critical density, as the latter is yet to be
determined 
at this stage. 
Secondly, from the fluid's coexistence curve expressed in terms of
the 
refractive index data, $n_{\rm L}$ and $n_{\rm V}$, the
critical refractive index, $n_{\rm C}$, is calculated from 
the coexistence curve diameter, 
$n_{\rm d}\equiv(n_{\rm L}+n_{\rm V})/2$. As $T\rightarrow T_{\rm
C}$, 
$n_{\rm d}\rightarrow n_{\rm C}$, in
accordance with the ``law of rectilinear diameter'', which has been
verified to hold for pure fluids~\cite{caitellet1886}, and it is 
also well obeyed by \dfe, as the data in Fig.~\ref{diameter} 
illustrate. The larger scatter in the data 
near the critical temperature is due to loss of accuracy in the 
measurements caused by high density gradients close to the critical 
point.
This value of $n_{\rm C}$ is then used in 
a plot of the \LLc, ${\cal L}\;vs.\;n$, to yield the critical value
of 
${\cal L}_{\rm C}={\cal L}(n_{\rm C})$.
Lastly, the values of $n_{\rm C}$ and ${\cal L}_{\rm C}$, and
Eq.~(\ref{LL}) give the critical density, 
$\rho_{\rm C}=(1/{\cal L}_{\rm C})(n_{\rm C}^{2}-1)/(n_{\rm
C}^{2}+2)$.

To have a reliable value of the critical temperature,
two correction--to--scaling terms must be retained in
Eq.~(\ref{corr2scaling}), while 
the critical temperature is treated as a free 
parameter in a nonlinear least square fit of the data. In this
experiment, we obtained
a value of $T_{\rm
C}=(302.964\pm 0.002)\,{\rm K}$ for the critical temperature of \dfe. 
The measured values for the critical refractive index and
critical \LLc\ are: $n_{\rm C}=1.1082\pm0.0006$, ${\cal L}_{\rm
C}=(0.1685\pm 0.0017)\;{\rm cm}^{3}\,{\rm g}^{-1}$, respectively.
We measured the critical density of \dfe\ as $\rho_{\rm
C}=(0.4195\pm0.0018){\rm g}\,{\rm cm}^{-3}$.

The coexistence curve results shown in 
Fig.~\ref{coex} display the typical trend of pure fluid, namely that
the coefficient $B_{1}$ for the first correction--to--scaling term of
Eq.~(\ref{corr2scaling}) is positive, 
as noticeable in the graph for $t>7\times 10^{-4}$. 

In one of our \LLc\ data sets (black circles in Fig.~\ref{LLdata}), 
we observed a slight increase at low densities
up to a value of $\rho\simeq 0.2\;\g\,\cm^{-3}$, where $\cal L$ has a
maximum.
The other set of data (dotted circles in Fig.~\ref{LLdata}) is
relatively constant in the same range. Both sets display 
a more marked monotonic decrease than previously observed in other 
fluids in the
rest of density interval explored, $(0.2<\rho<0.8)\;\g\,\cm^{-3}$.
We find no indication of
singular behavior in the vicinity of the critical density. These
features of the \LLc\ of \dfe\ are, to a certain extent, a departure from
similar measurements on other fluids. Previously, the \LL\ data either 
had a maximum at the critical density~\cite{burton74, ppmdb78} or did 
not display any strong monotonic trend~\cite{dbpp74}. Other
measurements showed a maximum for ${\cal L}(\rho)$ at a density below 
the critical density, but no strong monotonic decrease at higher
densities~\cite{uli90}.

Some experimenters have measured deviations from 
the \LLr. For example, Beysens~{\it et al.} measured discrepancies 
in the refractive index temperature coefficient at constant pressure,
$(\partial {\rm n}/\partial {\rm T})_{P}$~\cite{beysens77}. 
Others have indirectly 
determined values of the \LLc\ using their own refractive index 
measurements and tabulated values of density at known points, but 
their evaluations of the coefficient were carried out at only one point for 
each substance and are therefore of limited use in the search for any 
trends in the variation of the \LLc\ with density~\cite{chae90, schmidt94}.

We report in Table~\ref{critconst} a list of 
critical temperature and critical density measurements for \dfe\ found in the 
literature.
\begin{table*}
\begin{center}
    \begin{tabular}{lll} \hline\hline
	\multicolumn{1}{c}{Source} & \multicolumn{1}{c}{\hspace{3mm}
	$T_{\rm C}\,({\rm K})$} 
	& \multicolumn{1}{c}{\hspace{3mm}
	$\rho_{\rm C}\,({\rm g}\,{\rm cm}^{-3})$}\\\hline
	Ref.\cite{mears55} & \hspace{3mm}$303.25\pm 0.50$ & 
	\hspace{3mm}$0.417\pm 0.010$\\
	Ref.\cite{tsiklis67} & \hspace{3mm}$302.74\pm 0.005$ & 
	\hspace{3mm}$0.41\pm 0.02$\\
	Ref.\cite{CRC} & \hspace{3mm}$302.9$ & \hspace{3mm}$0.42$\\
	this article & \hspace{3mm}$302.964\pm 0.002$ & 
	\hspace{3mm}$0.4195\pm 0.0018$\\\hline\hline
    \end{tabular}
    \caption{Critical temperature and density of 
    1,1-difluoroethylene (C$_{2}$H$_{2}$F$_{2}$).}\label{critconst}
\end{center}    
\end{table*}
Our data agree well with these previously available results on the 
critical density and provide an improvement upon them. The slightly 
more marked discrepancies between our determination of the critical 
temperature and that found in these publications is likely due to 
different amounts and types of impurities in the 
samples~\cite{hastings80}.
We have been unable to find any published material on the \LLc\ of
research grade \dfe. Our experiments have therefore yielded the first 
accurate measurements of the \LLc\ of this material over a wide range 
of densities.
\vspace{10mm}

{\bf Acknowledgments}
\vspace{5mm}

We wish to thank Sharlene Tennant and Luisa Canuto 
for help in running the experiments. 
NF is also grateful to Douw Steyn 
for helpful input in the manuscript.

This research was funded in part by grants from the National Sciences
and
Engineering Research Council of Canada to DAB.

\end{document}